\documentclass[universe,article,accept,pdftex,moreauthors]{Definitions/mdpi}
\firstpage{1}
\pubvolume{1}
\issuenum{1}
\articlenumber{0}
\pubyear{2022}
\copyrightyear{2022}
\externaleditor{{Academic Editor:} 
}
\datereceived{15 June 2022}
\dateaccepted{15 July 2022}
\datepublished{}
\hreflink{https://doi.org/} 

\usepackage{aas_macros}
\usepackage{caption}
\usepackage{subcaption}
\usepackage{changes}
\usepackage{multirow}
\newcommand{\xadded}[1]{#1}
\newcommand{\xdeleted}[1]{}
\newcommand{\xreplaced}[2]{#1}

\Title{Determination of the Physical Parameters of AGNs in Seyfert 1 Galaxies LEDA 3095839 and VII Zw 244 Based on Spectropolarimetric Observations\\ \small{\it{Accepted for publication in Universe}}}

\TitleCitation{Determination of the Physical Parameters of AGNs in Seyfert 1 Galaxies LEDA 3095839 and VII Zw 244 Based on Spectropolarimetric Observations}

\Author{Elena Shablovinskaya $^1$, Mikhail Piotrovich $^{1,2,}$* \orcidA{}, Eugene Malygin $^1$, Stanislava Buliga $^{1,2}$ and \linebreak Tinatin Natsvlishvili $^2$}

\AuthorNames{Elena Shablovinskaya, Mikhail Piotrovich, Evgeny Malygin, Stanislava Buliga and Tinatin Natsvlishvili}

\AuthorCitation{Shablovinskaya, E.; Piotrovich, M.; Malygin, E.; Buliga, S.; Natsvlishvili, T.}

\address{%
$^{1}$ \quad Special Astrophysical Observatory RAS, 369167 Nizhnii Arkhyz, Russia; e.shablie@yandex.com (E.S.); male@sao.ru (E.M.); aynim@yandex.ru (S.B.)\\
$^{2}$ \quad Central Astronomical Observatory at Pulkovo RAS, 196140 Saint-Petersburg, Russia; tinatingao@mail.ru}
\corres{Correspondence: mpiotrovich@mail.ru}

\abstract{Here we present the detailed investigation of AGNs in two Seyfert 1 galaxies, LEDA 3095839 and VII Zw 244. Both of them were observed within the photometric reverberation mapping project in \xreplaced{Special Astrophysical Observatory of the Russian Academy of Sciences (SAO RAS) }{SAO RAS}, during which we earlier obtained the SMBHs masses. After that, both galaxies were observed in spectropolarimetric and polarimetric modes on the BTA 6 m telescope of the SAO RAS with the focal reducer SCORPIO-2. The linear polarization of the continuum and broad Balmer lines has been measured. It was found that (i) there were no signs of equatorial scattering in the LEDA 3095839 galaxy in the broad H$\alpha$ line, and we were able to estimate the value of SMBH spin and the magnetic field strength in the disk from the level of continuum polarization; (ii) for the galaxy VII Zw 244, the presence of equatorial scattering was shown, due to which the mass of the SMBH was independently measured, the inclination angle of the system was obtained, and the value of the spin was estimated.}


\begin{document}


\section{Introduction}

Active galactic nuclei (AGN), as the most powerful sources in the Universe, are unique objects where the physical state of the matter in extreme conditions can be tested. As~AGN radiation is produced due to the accretion of matter on the supermassive black holes (SMBH) in the center of host galaxies, the determination of such basic SMBH parameters as spin (dimensionless angular momentum) and mass is a crucial issue of modern astrophysics. Moreover, measuring these parameters is an important task, because~it allows us to trace the evolution of SMBH correlated with the evolution of galaxies at different cosmological epochs \citep{korho2013, heckman2014}.

\sloppy
\xreplaced{As for SMBH spin $a = c J / (G M_\text{BH}^2)$ (where $J$ is the angular momentum, $M_\text{BH}$ is the mass of the black hole, $G$ is the gravitational constant and $c$ is the speed of light), it is believed that the spin plays a central role in the generation of relativistic jets in AGN, and~the power of the relativistic jet is often used to determine the spin of the SMBH \citep{daly11}. It is believed that jets are formed due to the presence of a magnetic field in the disk. The~kinetic power of a relativistic jet can be obtained by estimating the magnetic field near the SMBH event horizon using several jet generation mechanisms: Blandford–Znajek mechanism \citep{blandford77}, Blandford–Payne mechanism \citep{blandford82} and Garofalo mechanism \citep{garofalo10}. }{As for SMBH spin $a = c J / (G M_\text{BH}^2$) (where $J$ is the angular momentum, $M_\text{BH}$ is the mass of the black hole, $G$ is the gravitational constant and $c$ is the speed of light), it has been established that the spin value plays a key role in the generation of relativistic jets in AGNs, therefore, it is the power of the relativistic jet that is most often used to determine the spin of the SMBH \citep{daly11}. It is believed that jets are formed due to the presence of a magnetic field in the disk. As~a rule, the~kinetic power of a relativistic jet is obtained by estimating the magnetic field strength near the SMBH event horizon using the Blandford–Znajek generation mechanism \citep{blandford77}. Other frequently used mechanisms are the Blandford–Payne mechanism \citep{blandford82}, and Garofalo mechanism \citep{garofalo10}.}

\xreplaced{Radiative efficiency $\varepsilon = L_{\rm bol} / (\dot{M} c^2$) (where $L_{\rm bol}$ is the bolometric luminosity of AGN and $\dot{M}$ is the accretion rate) depends significantly on the spin \mbox{\citep{bardeen72,novikov73,krolik07,krolik07b}}. So we could obtain the spin value by estimating the radiative efficiency. }{One of the effective methods for obtaining the spin value $a$ is to determine the radiative efficiency $\varepsilon(a)$ of the accretion disk, which depends significantly on the spin value of the black hole \mbox{\citep{bardeen72,novikov73,krolik07,krolik07b}}. Radiative efficiency is defined as $\varepsilon = L_{\rm bol} / (\dot{M} c^2$), where $L_{\rm bol}$ is the bolometric luminosity of AGN and $\dot{M}$ is the accretion rate.}

\xreplaced{The polarimetric observations of AGNs show polarization in different ranges \mbox{\citep{martin83,webb93,impey95,wilkes95,barth99,smith02,modjaz05,afanasiev11,afanasiev18}}. Spectropolarimetric characteristics of the accretion disk can demonstrate the effect of the presence of the magnetic field. Several mechanisms have been proposed for the occurrence of polarization in different structures (relativistic jets, the~plane or warped accretion disks, toroidal rings near the accretion disks), such as the light scattering in accretion disks or synchrotron radiation. There are several models of accretion disks (see for example \mbox{\citet{pariev03}}). The~most popular and simple model for our objects (Seyfert 1 AGNs with $0.01 < l_\text{E} < 0.3$ \mbox{\citep{netzer14}}, where $l_\text{E} = L_\text{bol}/L_\text{Edd}$ is Eddington ratio and $L_\text{Edd}$ is Eddington luminosity) is the optically thick, geometrically thin Shakura--Sunyaev disk \mbox{\citep{shakura73}}. }{The presence of a magnetic field should have a noticeable effect on the spectropolarimetric characteristics of the accretion disk radiation. The~polarimetric observations demonstrate that AGNs have polarized radiation in different wavelength ranges, from~ultraviolet to radio waves \mbox{\citep{martin83,webb93,impey95,wilkes95,barth99,smith02,modjaz05,afanasiev11,afanasiev18}}. Several mechanisms for the origin of the observed polarization are discussed, for~example, the~light scattering in accretion disks or synchrotron radiation of charged particles. These mechanisms can act in different structures, such as the plane and warped accretion disks, toroidal rings near the accretion disks, and relativistic jets. There are several models of accretion disks (see for example \mbox{\citet{pariev03}}). For~objects of the type under study, the~most popular {and simple} model is the optically thick geometrically thin Shakura--Sunyaev disk \mbox{\citep{shakura73}}.}

\fussy

The methods for determining masses of astrophysical objects are commonly based on the measurement of the matter rotating around. In~the case of SMBH, there are several direct and indirect approaches (see \cite{peterson2014} for a review), and~reverberation mapping is one of the most reliable direct methods used for low-redshift type 1 quasars \citep{blandford_revmap}. The~method is based on the estimation of the time delay between the continuum radiation of the accretion disk and the radiation in the broad emission lines produced in the BLR (broad-line region), which is why it requires a long observational time series. This raises an issue of measuring SMBH masses from a single-epoch observed spectrum of quasars \xreplaced{by calibration relations}{} which is still under development \citep{peterson2014}.

\citet{afanasiev15} \xreplaced{suggested }{established} a new alternative approach to trace the gas velocities in BLR in equatorially scattered type 1 AGN using single-epoch spectropolarimetric data. \xreplaced{The method is based on measuring the variation of the polarization angle as a function of BLR gas velocity upon the radiation of the differently rotating BLR disk reflects from the inner boundary of the dusty torus. A~detailed description is given further in Section~\ref{mass_est}}{} This method is still applied to a relatively small number of close Seyfert 1 galaxies ($\sim$30 AGN) (see \cite{afanasiev19,savic2021}), yet it revealed the statistically approved and stable results of SMBH \xreplaced{measurements}{measuring}. Moreover, in~the same work it was shown that comparison between two independent SMBH mass measurements given by the spectropolarimetric and reverberation mapping approaches \xreplaced{allows}{ allow} one to estimate the inclination angle of the AGN relative to the~observer.

\xreplaced{Therefore, as~it was described above, the~polarimetric tools provide us with valuable information about the structure and physical parameters of geometrically unresolved central parts of AGN. This allows us to take a more detailed look at the physical characteristics of SMBH and gas around in individual AGNs to build more general models of the galaxy's evolution. For~a thorough analysis of the physical parameters previously (see \cite{afanasiev19}) for conducting spectropolarimetric observations on the 6 m BTA telescope and searching for signs of equatorial scattering, the~main attention was paid to the sample of AGNs with SMBH masses measured by reverberation mapping approach (e.g., \cite{bentz13}). Within~this work we decided to focus our attention on the AGNs examined earlier by our group in the Special Astrophysical Observatory of the Russian Academy of Sciences (SAO RAS) using the photometric reverberation mapping \citep{Uklein19}. Currently, there are two type 1 AGN from that sample---LEDA 3095839 and VII Zw 244, where we have estimated the BLR size $R_{\rm BLR}$ and, consequently, the~SMBH masses (or so-called virial products, see Section~\ref{ffact}) \citep{malygin20}. It motivated us to select these objects for further investigation. The~polarimetric observations revealed the non-zero level of the polarization in spectra of both galaxies, yet a further detailed examination showed the diverse origin of the observed effects. }{Therefore, as~it was described above, the~polarimetric tools provide us with valuable information about the structure and physical parameters of geometrically unresolved central parts of AGN. The~AGN polarimetric and especially spectropolarimetric observations are intensively being conducted on the 6 m BTA telescope at SAO RAS. Within~this work we provide and analyze the spectropolarimetric data obtained for two type 1 AGN, LEDA 3095839 and VII Zw 244. These two galaxies were examined with the photometric reverberation mapping provided earlier by our group in SAO \citep{Uklein19}. Currently, for the given objects we have estimated the BLR size $R_{\rm BLR}$ and, consequently, the~SMBH masses (or so-called virial products, see Section~\ref{inclinZw}) \citep{malygin20}. The~polarimetric observations revealed the non-zero level of the polarization in spectra of both galaxies, yet a further detailed examination showed the diverse origin of the observed effects. }

The paper is organized as follows. In~Section~\ref{sec2} we describe the polarimetric observations and data reduction procedures. The~results of the data analysis \xreplaced{are}{is} provided in Section~\ref{sec3}, where for the objects the continuum or broad line polarization is estimated, and bolometric luminosity, inclination angle, spin, and magnetic field strength are calculated. The~conclusions are given in Section~\ref{sec4}.


\section{Observations and~Reduction}\label{sec2}

The observations were carried out with the 6-meter BTA telescope of the SAO RAS. We used a modified SCORPIO-2 spectrograph \citep{afanasiev11b} in spectropolarimetric and polarimetric modes.
The observation log is presented in Table~\ref{tab_obs}, which shows the object names, redshift $z$, observational mode, volume phase holographic grism (VPHG, for~spectropolarimetry) or medium-band filter from SED-set\endnote {Details about the characteristics of medium-band filters can be found at \url{https://www.sao.ru/hq/lsfvo/devices/scorpio-2/filters_eng.html} Accessed 15 Jul. 2022.} (for polarimetry), date of observations, exposure time, the~direct image quality (seeing), air mass, and position angle (PA). For~both polarimetry and spectropolarimetry we use a double Wollaston prism \citep{oliva97,geyer1993} as a polarization analyzer, that allows one to simultaneously obtain four images corresponding to the electric-vector oscillation directions of 0$^\circ$, 90$^\circ$, 45$^\circ$, and 135$^\circ$ and, consequently, three Stokes parameters $I$, $Q$, and~$U$ within one~exposure.

It is important to note that on 18 November 2019, during~polarimetric observations of LEDA 3095839, the~zenith distance of the Moon illuminated by 60\% was $\sim$38$^\circ$. On\linebreak 12 October 2020, during~the spectropolarimetric observations of VII Zw 244, the~Moon, illuminated by 19\%, rose to a zenith distance of 74$^\circ$. At~the same time, the~obtained polarization values coincided with the measurements of the same objects observed on moonless~nights.

\begin{table}[H]
\caption{Log of BTA observations of studied~AGN.}
\label{tab_obs}
\newcolumntype{C}{>{\centering\arraybackslash}X}
\begin{adjustwidth}{-\extralength}{0cm}
\centering 
\begin{tabularx}{\fulllength}{CCCCCCCCC}
\toprule
\textbf{Object}                        & \boldmath$z$                      & \textbf{Mode}    & \begin{tabular}[c]{@{}c@{}}\textbf{VPHG}\\/\textbf{Filter}\end{tabular}          & \begin{tabular}[c]{@{}c@{}}\textbf{Date}\\ \textbf{dd/mm/yy}\end{tabular} & \begin{tabular}[c]{@{}c@{}}\textbf{Exposure,}\\ \textbf{s}\end{tabular}                                                       & \textbf{Seeing} & \textbf{Air Mass} & \begin{tabular}[c]{@{}c@{}}\textbf{PA,}\\ \textbf{Degrees}\end{tabular}    \\ \midrule
\multirow{2}{*}{LEDA 3095839} & \multirow{2}{*}{0.109} & specpol & 1026@735                                                         & 3 March 20                                                & 300 + 9 $\times$ 600                                                       & 1".7   & 1.4      & 106.3 \\ \cmidrule{3-9}
                              &                        & pol     & \begin{tabular}[c]{@{}c@{}}SED700\\ SED725\end{tabular}          & 18 November 2019                                                & \begin{tabular}[c]{@{}c@{}}60 + 4 $\times$ 80\\ 60 + 4 $\times$ 50\end{tabular}   & 1".3   & 1.2      & 55.6  \\ \midrule
\multirow{2}{*}{VII Zw 244}   & \multirow{2}{*}{0.131} & specpol & 940@600                                                          & 12 October 2020                                                & 6 $\times$ 900                                                             & 1".6   & 1.3      & 0.0     \\ \cmidrule{3-9}
                              &                        & pol     & \begin{tabular}[c]{@{}c@{}}SED700\\ SED725\\ SED750\end{tabular} & 7 November 2019                                                & \begin{tabular}[c]{@{}c@{}}7 $\times$ 180\\ 7 $\times$ 180\\ 7 $\times$ 180\end{tabular} & 2".5   & 1.3      & 54.2  \\ \bottomrule
\end{tabularx}
\end{adjustwidth}
\end{table}

\subsection{Spectropolarimetry}

In the spectropolarimetric mode, the~slit length is 1'. In~the case of observations of both AGNs, the~slit width was $\sim$2''. The~spectral resolution is determined by the slit width and the used spectral grating\endnote{Details about the characteristics of gratings can be found at \url{https://www.sao.ru/hq/lsfvo/devices/scorpio-2/grisms_eng.html} Accessed 15 Jul. 2022.}. Spectropolarimetric data for LEDA 3095839 was obtained on March 3, 2020 with the total exposure of $\sim$1.6 h. The~VPHG1026@735 grism with the CCD detector E2V 42-90 (with a pixel size of 13.5$\times$13.5 $\upmu$m; in spectral mode with binning \linebreak 2 $\times$ 1, giving a frame of size $\sim$2 k $\times$ 2 k px and scale 0''.352 $\times$ 0''.176 per pixel$^2$) provided the spectrum coverage of 5800--9500\AA\ with the FWHM $\sim$10\AA. In~this configuration, a~blocking GS-17 filter was also used for object LEDA 3095839 to suppress the grating second order. We also note that for LEDA 3095839, we do not provide here the red part of the spectrum beyond 8200\AA\ due to the presence of "fringes"\ for CCD E2V 42-90, which leads to large noise upon subtracting the sky background. The~BTA prime focus adapter was used to calibrate the spectrum \citep{adapter}. For~this, we obtained the comparison spectrum from a He-Ne-Ar lamp and flat field frames. An~extremely important calibration of the three-dot mask (three-dot test) was also obtained, which makes it possible to take into account the geometric distortions of the field in received frames, which is critical for the analysis of the polarization of such faint objects as AGNs.

Spectropolarimetric data for VII Zw 244 was obtained on 12 October 2020. The~total exposure was 1.5 h. VPHG940@600 with 2$"$-wide slit, with~LP425 filter (for second order cutoff) together with CCD E2V 261-84 (with pixel size 15 $\times$ 15 $\upmu$m; in spectral mode with \linebreak 2 $\times$ 4 binning, giving a frame size of $\sim$2 k $\times$ 520 px and scale 0$''$.391 $\times$ 0$''$.782 per \mbox{pixel$^2$ \citep{popovic2021})} provided the spectrum coverage of 3500--8500\AA\ with FWHM $\sim$12\AA. The~comparison spectrum, three-dot test, and flat fields were also taken with the prime focus adapter. Observations of objects were accompanied by observations of polarized standard stars and stars with zero polarization.

Introducing the instrumental parameters $K_Q$ and $K_U$, which characterize the transmission of polarization channels, determined from observations of unpolarized standard stars, as~well as $I_{0}$($\lambda$), $I_{45}$($\lambda$), $I_{90}$($\lambda$), and $I_{135}$($\lambda$) as the intensity at four polarization directions, we can measure three Stokes parameters:
\begin{equation}
\label{formula:stokes_I}
I(\lambda)=I_{0}(\lambda) + I_{90}(\lambda)K_{Q}(\lambda) + I_{45}(\lambda) + I_{135}(\lambda)K_{U}(\lambda)
\end{equation}
\begin{equation}
\label{formula:stokes_q}
Q(\lambda) =\frac{ I_{0}(\lambda) - I_{90}(\lambda)K_{Q}(\lambda) }{ I_{0}(\lambda) + I_{90}(\lambda)K_{Q}(\lambda) } \end{equation}
\begin{equation}
\label{formula:stokes_u}
U(\lambda) =\frac{ I_{45}(\lambda) - I_{135}(\lambda)K_{U}(\lambda) }{ I_{45}(\lambda) + I_{135}(\lambda)K_{U}(\lambda) }
\end{equation}

 Here and below we use $Q$ and $U$ to denote the normalized Stokes~parameters.

Then, by~determining the zero-point of the polarization angle $\varphi_{0}$ from observations of polarized standard stars, one can determine the degree of linear polarization $P(\lambda)$ and the polarization angle $\varphi(\lambda)$ as:
\begin{equation}
\label{formula:stepen}
P(\lambda)= \sqrt{Q^{2}(\lambda)+U^{2}(\lambda)}
\end{equation}
\begin{equation}
\label{formula:ugol}
\varphi(\lambda) =\frac{1}{2} \arctan[\frac{ U(\lambda) }{ Q(\lambda) }] + \varphi_{0}
\end{equation}

The observation technique and data reduction are described in more detail in \citep{afanasiev12}. Interstellar medium (ISM) polarization and atmospheric depolarization were taken into account using polarimetric measurements of field stars around the studied AGN.

\subsection{Polarimetry}

In the polarimetric mode with double Wollaston prism for LEDA 3095839, we obtained 5 frames each with $\sim$1 minute exposure time in interference SED filters with bandwidths of the order of $\sim$250 \AA\ with 2 $\times$ 2 binned (frame size 1 k $\times$ 1 k px and scale 0$"$.352 $\times$ 0".352 per pixel) CCD E2V 42-90. One filter is oriented to the H$\alpha$ emission line, and~the second one is oriented to the continuum near. For~VII Zw 244, we obtained polarimetric data using the same CCD with the same binning in three filters (one covers the H$\alpha$ line, and~two cover the continuum around), 7 frames each with $\sim$3 min exposure time. \xreplaced{In case of observations of both objects }{} in the image polarimetry mode, the~double Wollaston prism produces four images of the entrance pupil bounded by a rectangular mask, so each image has a field of view of $\sim$6' $\times$1'.

It is important to clarify that in both cases we oriented the instrument so that the masked field contained bright non-variable stars (at a distance of 1$'$.6 from LEDA 3095839 and at a distance of 4'.6 from VII Zw 244). \xreplaced{This allows the use of a differential polarimetry technique that minimizes the effects of the depolarization due to the atmospheric variations that can reach up to 2--3\% \citep{afanasiev12} and instrumental errors mostly due to the prism transmission which is about 0.2\% according to the independent measurements of the polarized standard stars \citep{afanasiev19}. }{This allows the use of a differential polarimetry technique that minimizes systematic and instrumental errors associated with variable atmospheric depolarization and prism transmission.}  To find \xreplaced{more }{more more} about differential polarimetry see \citep{shablovinskaya19}. This technique ensured an accuracy of polarimetry of about $\sim$0.1\% for the field stars. Within~this work, obtaining AGN polarimetry data in filters allows us to refine the average polarization value in objects and roughly estimate the effect of ISM~polarization.

Since the type 1 AGN linear polarization is usually close to zero, as~it was shown \linebreak in \citep{simmons1985}, when calculated by \xreplaced{Equation }{formula} (\ref{formula:stepen}), the~degree of polarization $P$ turns out to be biased. So, for~spectropolarimetric\xreplaced{}{and polarimetric} observations, we used the following relation to obtain the unbiased value of the polarization degree:
\xreplaced{\begin{equation}
P_{\rm unbiased} = P \cdot \sqrt{1 - (K\cdot\sigma_{P}/P)^{2}},
\label{formula:unbias}
\end{equation}}{$
P_{\rm unbiased} = \sqrt{P^{2} - 1.41\cdot\sigma_{P}^{2}}$}
where $P$ is \xreplaced{measured degree of fractional linear polarization }{measured polarization degree and}, $\sigma_{P}$ is the corresponding error \xreplaced{obtained from observational data as the standard deviation, $\sqrt{1 - (K\cdot\sigma_{P}/P)^{2}}$ is called a Rice factor. \citet{simmons1985} recommended the bias correction factor $K=1.41$ for low signal-to-noise ratios ($\sigma_{P}/P \lesssim 0.7$) based on the maximum likelihood estimator of the true value of the polarization degree.}{} \xreplaced{Note that we do not correct the measured polarization values obtained using image polarimetry, because~the measured accuracy is an order of magnitude higher than the accuracy of  spectropolarimetric measurements. }{}

\section{Results}\label{sec3}
\label{results}

\subsection{Object LEDA~3095839}

\subsubsection{Continuum~Polarization}

\begin{figure}[H]
    \includegraphics[scale=0.5]{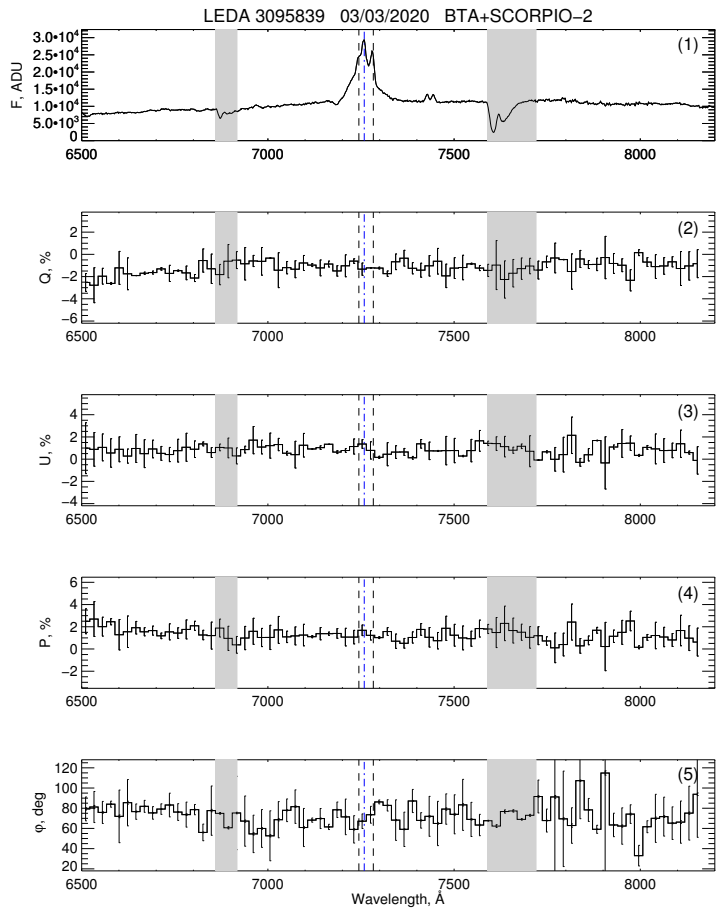}
    \caption{Spectropolarimetric data for LEDA 3095839 in the range of 6500--8200 \AA. From~the top to the bottom:  integral spectrum in ADU $F$ (\textbf{1}), normalized Stokes parameters $Q$ (\textbf{2}) and $U$ (\textbf{3}), polarization degree $P$ (\textbf{4}), and polarization angle $\varphi$ (\textbf{5}). The~vertical blue dash-dot line denote the position of H$\alpha$ line; the vertical black dash lines denote the position of narrow [NII] lines. The~atmospheric absorption bands are noted with light grey~stripes. }
    \label{fig:sppolLEDA}
\end{figure}

The results of LEDA 3095839 spectropolarimetry are shown in Figure~\ref{fig:sppolLEDA}. The~spectrum is given in the 6500--8200 \AA\ wavelength range, where the shifted broad H$\alpha$ line is~located.

The first panel in Figure~\ref{fig:sppolLEDA} shows the integral spectrum in ADU units, not corrected for the device sensitivity curve. Note that the atmospheric absorption bands of molecular oxygen B-band (6860--6917 \AA) and A-band (7590--7720 \AA) are visible in the continuum spectrum of the object; in addition, in~the range $\sim$7150--7350 \AA\ there is a water vapor absorption band, which appears exactly in H$\alpha$ broad line. On~panel 1 and others, the~H$\alpha$ line center and the [NII] narrow lines are~marked.

The second and third panels in Figure~\ref{fig:sppolLEDA} show the Stokes parameters $Q$ and $U$ binned over the 22 \AA-window, corrected for the Wollaston prism transmission and not corrected for the rotation angle of the device. The~step of the spectral bin was chosen so that, while estimating the robust average of the measured values over the spectral range and all taken exposures, the~outlier points (mostly due to the cosmic ray hints) were corrected. The~errors given on the plots are equal to the 1$\sigma$ level, where $\sigma$ is the robust standard deviation. \xreplaced{Here for the analysis we use the basic robust estimation using 2$\sigma$ rejection threshold. More details about the algorithm can be found in \cite{numres} and references therein}. It can be seen that within the broad H$\alpha$ line, there are no significant changes in the $Q$ and $U$ parameters along wavelength, and~their variations do not exceed the measurement error. This is the main marker of the absence of the equatorial scattering in the object. Thus, the~average value of the Stokes parameters over the observed spectrum is $<Q> = -(1.2\pm 0.4)$\% and $<U> = (0.8\pm 0.4)$\%. These values coincided with the averaged values from the observations in the polarimetric mode in the medium-band filters on another night:
$Q_{\rm f} = -(1.0 \pm 0.1)$\% and $U_{\rm f} = (0.8 \pm 0.2)$\%, which indicates the stability of the object's polarization state on the timeline of~observations.

Polarization degree $P$ (panel 4 in Figure~\ref{fig:sppolLEDA}) also shows no deviations in the H$\alpha$ line from the mean value of the continuum polarization. At~the same time, it can be seen that the polarization level of the object is non-zero, does not change along the wavelength within the error and, \xreplaced{upon }{with} robust averaging over the entire observed spectrum \xreplaced{(which is almost equal to the median value of $P$)}{}, is $<P> = (1.3 \pm 0.4)$\%. Polarization angle $\varphi$ (panel 5 in Figure~\ref{fig:sppolLEDA}) also does not demonstrate dependence on the wavelength and, in~particular, features along the broad line. So, it can also be considered constant within the error with a robust average value of $<\varphi> = (73 \pm 8)^\circ$. Note that the presented polarization angle $\varphi$ is given in instrumental units that are not rotated to the north direction on the celestial plane. The~obtained estimates also coincide with the polarization estimates in the filters: $P_{\rm f} = (1.3\pm 0.2)$\% and $\varphi_{\rm f}= (71 \pm 4)^\circ$.

Since the galactic latitude of the object is $b\sim +33^\circ$, it is worth assuming that the observed polarization is partially or completely influenced by ISM polarization together with depolarization in the atmosphere and the influence of residual instrumental effects. Observations in filters in the polarimetry mode showed that the object's environment is weakly polarized, however, the~average polarization value of the neighboring three stars in the field is determined with a large error and is $P_{\rm stars}\sim (2\pm1)$\% and $\varphi_{\rm stars}\sim (80\pm15)^{\circ}$. This shows that a significant part of the observed polarization level in the object does not relate to the nature of the source. However, if~one tries to correct the additional polarization by the star closest to the object\endnote{The GAIA parallax of the star (2MASS J08540279+7700478) indicates it is located at the distance of 400 pc \citep{gaiadr3}.}, then even in this case, the minimum possible intrinsic polarization of the object is $P \sim 0.5$\%. So far, it appears that the observed continuum polarization of the source cannot be totally caused by ISM polarization. \xreplaced{It is also important to mention that in the given case in Figure~\ref{fig:sppolLEDA} the values of $P$ are presented biased because they could not be properly corrected due to the field polarization providing non-zero $Q$ and $U$ Stokes parameters. }{}

Since we have not found any signs of equatorial scattering in LEDA 3095839 spectra, we further assume that the observed polarization related to the source is generated in the accretion disk. Due to the large error in determining the correction for ISM polarization, we believe that the continuum polarization level in the object is between 0.5\% and 1.3\%. Thus, further, to~determine the physical parameters of AGN, we assume for LEDA 3095839 $P = (0.9\pm0.4)$\%. However, it is worth mentioning that if the object intrinsic polarization is higher, up~to 1.3\%, then it is worth assuming not only polarization mechanisms in the accretion disk, but~the complex influence of polarized radiation from different AGN structures, e.g.,~synchrotron radiation of a jet. Within~the framework of this paper, we neglect the influence of other polarization~mechanisms.

\subsubsection{Bolometric~Luminosity}
\sloppy
According to \citet{xu07}, the bolometric luminosity of LEDA 3095839 is \linebreak $\log{[L_\text{bol}(erg/s)]} = 45.05$. They derive this value from luminosity in $B$ filter \linebreak $\log{[L_\text{B}(erg/s)]} = 44.39$ using bolometric correction \citep{xu03}. Our observations on the 6 m BTA telescope give a luminosity value at 5100\AA\, $\log{[L_\text{5100}(erg/s)]} = 43.94$ \citep{malygin20}. There is a serious problem in defining the bolometric correction factors that are used for obtaining luminosity value at a certain wavelength from bolometric luminosity. In~the literature, we can find factors that differ 2--3 times \citep{richards06, hopkins07, cheng19, netzer19, duras20}. In~this work we decided to use bolometric correction for luminosity at 5100 \AA\, from~\citet{richards06}: $L_\text{5100} = L_\text{bol} / 10.3$. Thus we get the value $\log{[L_\text{bol}(erg/s)]} = 44.95$.
\fussy

\subsubsection{Inclination Angle and Spin~Value}
\label{ffact}

To determine the spin value $a$ we need to obtain radiative efficiency $\varepsilon$ \xadded{which depends significantly on the spin \citep{bardeen72,novikov73,krolik07,krolik07b}. Then we could determine spin numerically using the relation \citep{bardeen72}:}

\xadded{
\begin{equation}
 \varepsilon(a) = 1 - \frac{R_\text{ISCO}^{3/2} - 2 R_\text{ISCO}^{1/2} + |a|}{R_\text{ISCO}^{3/4}(R_\text{ISCO}^{3/2} - 3 R_\text{ISCO}^{1/2} + 2 |a|)^{1/2}},
 \label{eq01}
\end{equation}}

\noindent \xadded{where $R_\text{ISCO}$ is the radius of the innermost stable circular orbit of a black hole and}

\xadded{
\begin{equation}
 \begin{array}{l}
  R_\text{ISCO}(a) = 3 + Z_2 \pm [(3 - Z_1)(3 + Z_1 + 2 Z_2)]^{1/2}\\
  Z_1 = 1 + (1 - a^2)^{1/3}[(1 + a)^{1/3} + (1 - a)^{1/3}]\\
  Z_2 = (3 a^2 + Z_1^2)^{1/2}
 \end{array}
\end{equation}}

 \xadded{Here the sign ''-'' is used for prograde ($a \geq 0$), and~the sign ''+'' for retrograde \linebreak rotation ($a < 0$).}

There are several models connecting the radiative efficiency with such parameters of AGN as mass of SMBH $M_\text{BH}$, angle between the line of sight and the axis of the accretion disk $i$ and bolometric luminosity $L_\text{bol}$ 
\citep{davis11,raimundo11,du14,trakhtenbrot14,lawther17}.
Based on the results of our previous \mbox{work \citep{piotrovich22}} we decide to use \xadded{theoretical} model \xadded{developed for Shakura--Sunyaev accretion disk case} from \citet{du14}:
\begin{equation}
  \varepsilon \left( a \right) =  0.105 \left(\frac{L_\text{bol}}{10^{46}\text{erg/s}}\right) \left(\frac{L_{5100}}{10^{45}\text{erg/s}} \right)^{-1.5} M_8 \mu^{1.5}
\end{equation}

 Here $M_8 = M_\text{BH} / (10^8 M_{\odot})$, $\mu = \cos{(i)}$ and $i$ is the inclination angle between the line of sight and the axis of the~disk.

We used the SMBH mass value from \citet{malygin20} calculated for $f=1$, where $f$ is the dimensionless parameter characterizing the object geometry  \xreplaced{and orientation that is usually introduced as following (see \cite{peterson2014}):
\begin{equation}
    M_{\rm BH} = f \times (R_{\rm BLR}\vartheta^2_\text{line}G^{-1}) = f \times V.P.,
    \label{vp}
\end{equation}
where $\vartheta_\text{line}$ is the velocity of the line-emitting gas in the BLR, $G$ is the gravitational constant, and~$V.P.$ is a virial product, which characterizes the least possible value of the $M_{\rm BH}$ as $f\geq 1$ ($f \approx 1 /[\sin^2{(i)}]$ due to the projection of the gas velocity $\vartheta_\text{line}$ to the line of sight). }{(in this case, the~result is so-called "virial product", see Section~\ref{inclinZw}.)} \xreplaced{So far }{Taking into account the fact that $M_\text{BH} \sim f$ and $f \approx 1 /[\sin^2{(i)}]$}, we assumed that polarized radiation from this object is produced by accretion disk. Thus, based on our polarimetric data [$P = (0.9\pm0.4)$\%] and Sobolev--Chandrasekhar mechanism \citep{sobolev63,chandrasekhar50,gnedin15} with addition of Faraday depolarization from magnetic field \citep{piotrovich21} we concluded that inclination angle must satisfy the condition $35^{\circ} \lesssim i \lesssim 55^{\circ}$. Next, we considered three options: $i = 35^\circ$, $i = 45^\circ$, and $i = 55^\circ$. Table~\ref{tab01} shows results of our calculations for first two options. The~results are quite typical for objects of this type \citep{piotrovich22}. Results for $i = 55^\circ$ are inconclusive and do not allow to determine the value of spin, so we decided not to consider this option~further.

\renewcommand{\arraystretch}{1.5}
\begin{table}
    \setlength{\tabcolsep}{3pt}
    \caption{Results of our calculations of masses $M_\text{BH}$ and spins $a$ for our objects. $P$ is polarization degree in [\%], $FWHM$ is full width at half maximum of H$\beta$ line in km/s, $L_\text{bol}$ is bolometric luminosity in erg/s, $l_{\text{E}}$ is Eddington ratio, $i$ is inclination angle in degrees, $\varepsilon$ is radiative efficiency, $B_\text{H}$ and $B^*_\text{H}$ are magnetic field strength at event horizon in Gauss obtained from spectral characteristics and from polarization data respectively, $s$ is the exponent of the power-law dependence of the magnetic field on the radius. For VII Zw 244, the SMBH mass and inclination angle estimations obtained by spectropolarimetry are marked with $^\dagger$ symbol.}
    \label{tab01}
    \newcolumntype{C}{>{\centering\scriptsize\arraybackslash}X}
    \begin{adjustwidth}{-\extralength}{0cm}
    \centering
    \begin{tabularx}{\fulllength}{CCCCCCCCCCCCC}
    \toprule
    \textbf{Object} & \boldmath$P$ & \boldmath$FWHM$ & \boldmath$\log(L_\text{bol})$ & \boldmath$l_\text{E}$ & \boldmath$i$ & \boldmath$\log(M_\text{BH}/M_{\odot})$ & \boldmath$\varepsilon$ & \boldmath$a$ & \boldmath$\log(B_\text{H})$ & \boldmath$\log(B^*_\text{H})$ & \boldmath$s$ \\
    \hline
    \multirow{2}{*}{LEDA 3095839} & \multirow{2}{*}{$0.9\pm0.4$} & \multirow{2}{*}{3775} & \multirow{2}{*}{44.95} & $0.11^{+0.09}_{-0.05}$ & 35 & $7.881^{+0.153}_{-0.171}$ & $0.21^{+0.09}_{-0.07}$ & $0.966^{+0.030}_{-0.106}$ & $4.06^{+0.24}_{-0.24}$ & $3.53^{+0.26}_{-0.53}$ & 1.77$\pm$0.18 \\
                                  &                              &                       &                        & $0.22^{+0.18}_{-0.09}$ & 45 & $7.699^{+0.153}_{-0.171}$ & $0.11^{+0.10}_{-0.04}$ & $0.736^{+0.226}_{-0.368}$ & $4.00^{+0.56}_{-0.34}$ & $4.08^{+0.30}_{-1.08}$ & 1.63$\pm$0.23 \\
    \hline
        \multirow{2}{*}{VII Zw 244} & \multirow{2}{*}{$1.4\pm0.6$} & \multirow{2}{*}{3219} & \multirow{2}{*}{45.24} & $0.11^{+0.03}_{-0.02}$ & 18 & $8.069^{+0.068}_{-0.075}$ & $0.30^{+0.02}_{-0.06}$ & $0.996^{+0.002}_{-0.012}$ & $4.29^{+0.10}_{-0.13}$ & -- & -- \\
                                    &                              &                       &                        &                        & 14.3$\pm$3.6$^{\dagger}$ & $8.29\pm0.30^{\dagger}$ & -- & -- & -- & -- & -- \\
    \bottomrule
    \end{tabularx}
    \end{adjustwidth}
\end{table}
\renewcommand{\arraystretch}{1.00}

\subsubsection{Magnetic~Field}

\xreplaced{We were unable to find information about the radio properties of the LEDA 3095839 object in the literature. So we operated on the assumption that it is radio loud and has a jet. Thus, we can estimate the magnetic field strength using the approach from \mbox{\citet{piotrovich20}}: }{Using the approach from \mbox{\citet{piotrovich20}} we can estimate the magnetic field strength with this expression:}
\begin{equation}
 B_H = \frac{10^{(5.78 \pm 0.07)} \eta \sqrt{\varepsilon}}{l_E^{(0.295 \pm 0.020)} |a| [\cos{(i)}]^{3/4}} \left(\frac{10^3 \text{km/s}}{FWHM}\right)^3 \text{G}
 \label{eq02}
\end{equation}

Here $\eta$ is the coefficient depending on the model of relativistic jet generation (in this case $\eta = 1.05^{-0.5}$ \citep{piotrovich20}) and $FWHM$ is full width at half maximum of spectral line H$\beta$. We use the Eddington ratio $l_\text{E} = L_\text{bol} / L_\text{Edd}$, where $L_\text{Edd} = 1.5 \times 10^{38} M_\text{BH} / M_\odot $ is the Eddington luminosity. For~this object we take value $\vartheta_\text{line} = 1.5 \times 10^3$ km/s from \citet{malygin20} and calculate $FWHM$ = 2.335 $k^{-0.5} \vartheta_\text{line}$. Here $k \approx 0.88$ is the coefficient arising from the fact that in \citet{malygin20} $\vartheta_\text{line}$ for this object referred to H$\alpha$ line \citep{Uklein19}.

Additionally, we use the different approach from \citet{piotrovich21} that relates the geometry of the magnetic field in an accretion disk to the polarization of its radiation. So we can estimate the magnetic field strength at the event horizon of SMBH $B^*_\text{H}$ and the exponent $s$ of the power-law dependence of the magnetic field on the radius $B^*(R) = B^*_\text{H} (R_\text{H} / R)^s$, where $R_\text{H} = G M_\text{BH} (1 + \sqrt{1 - a^2}) / c^2$ is the radius of the event~horizon.

The results of both methods are shown in Table~\ref{tab01}. Obtained values of the magnetic field are sufficiently close and quite typical for objects of this type \citep{piotrovich21,daly19}. A~relatively large error in determining the magnetic field is associated with significant errors in the measurement of the spin value and relatively small polarization~degree.

\subsection{Object VII Zw~244}

\subsubsection{Polarization in Broad~Lines}

\begin{figure}[H]
    \includegraphics[scale=0.6]{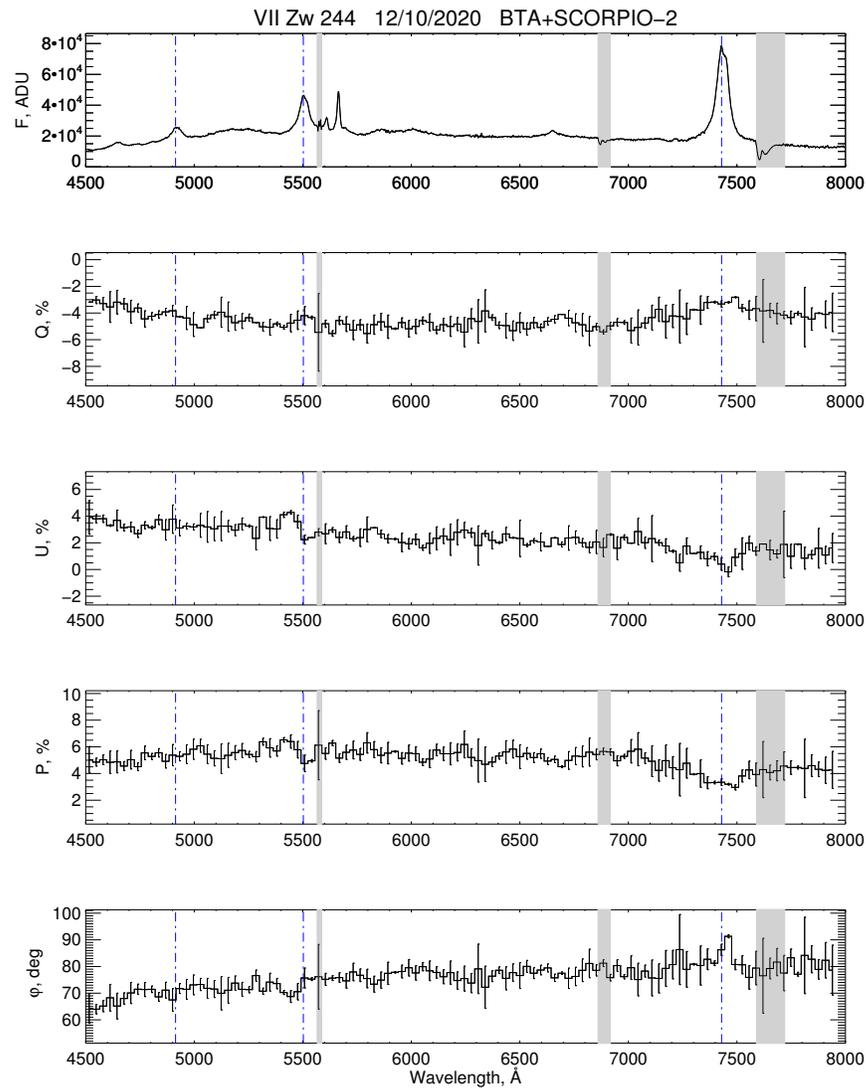}
    \caption{Spectropolarimetric data for VII Zw 244 in the range of 4500-8000\AA. From~the top to the bottom:  integral spectrum in ADU $F$ (\textbf{1}), normalized Stokes parameters $Q$ (\textbf{2}) and $U$ (\textbf{3}), polarization degree $P$ (\textbf{4}), and polarization angle $\varphi$ (\textbf{5}). The~vertical blue dash-dot line denote the position of (from the left to the right) H$\gamma$, H$\beta$, and H$\alpha$ lines. The~atmospheric absorption/emission bands are noted with light grey~stripes. }
    \label{fig:sppolZw}
\end{figure}

The results of AGN VII Zw 244 spectropolarimetry are shown in Figure~\ref{fig:sppolZw}. The~spectrum is given in the wavelength range 4500--8000\AA\ and contains shifted broad lines H$\alpha$, H$\beta$, and H$\gamma$.
The panels on Figure~\ref{fig:sppolZw} correspond to the panels on Figure~\ref{fig:sppolLEDA}, so the details will be omitted here. The~atmospheric absorption bands of molecular oxygen B-band, A-band, and the molecular oxygen emission line 5577 \AA\ are marked on the~spectra.

The second and third panels on Figure~\ref{fig:sppolZw} show the Stokes parameters $Q$ and $U$ binned over the 32 \AA-window, corrected for the Wollaston prism transmission and not corrected for the rotation angle of the device. The~degree $P$ and the polarization angle $\varphi$ are binned in the same window. \xreplaced{Here the observed values of $P$ not corrected to the bias are presented.}{}

In the entire observed range, the~Stokes parameters $Q$ and $U$ show both a slight smooth dependence on the wavelength \xreplaced{(mostly in the case of $U$)}{}, and~the dramatic variations within the broad lines, \xreplaced{where opposite variation patters could be detected in $Q$ and $U$ parameters}{}. It is important to note that the pattern of the parameter changes in the two lines H$\alpha$ and H$\beta$ repeats each other up to noise; in the H$\gamma$ line, the effect is too small and does not exceed noise, which is why further analysis of polarization is not carried out in this line. It is clearly seen that the degree of polarization of $P$ in broad lines differs significantly (by $\sim$1--1.5\%) from the continuum value, and~the angle acquires an S-shaped profile with an amplitude significantly exceeding the measurement error. These features indicate the presence of equatorial scattering in the central regions of the AGN \citep{smith05}.

\begin{figure}[H]
     \begin{subfigure}[b]{0.4\textwidth}
         \includegraphics[width=\textwidth]{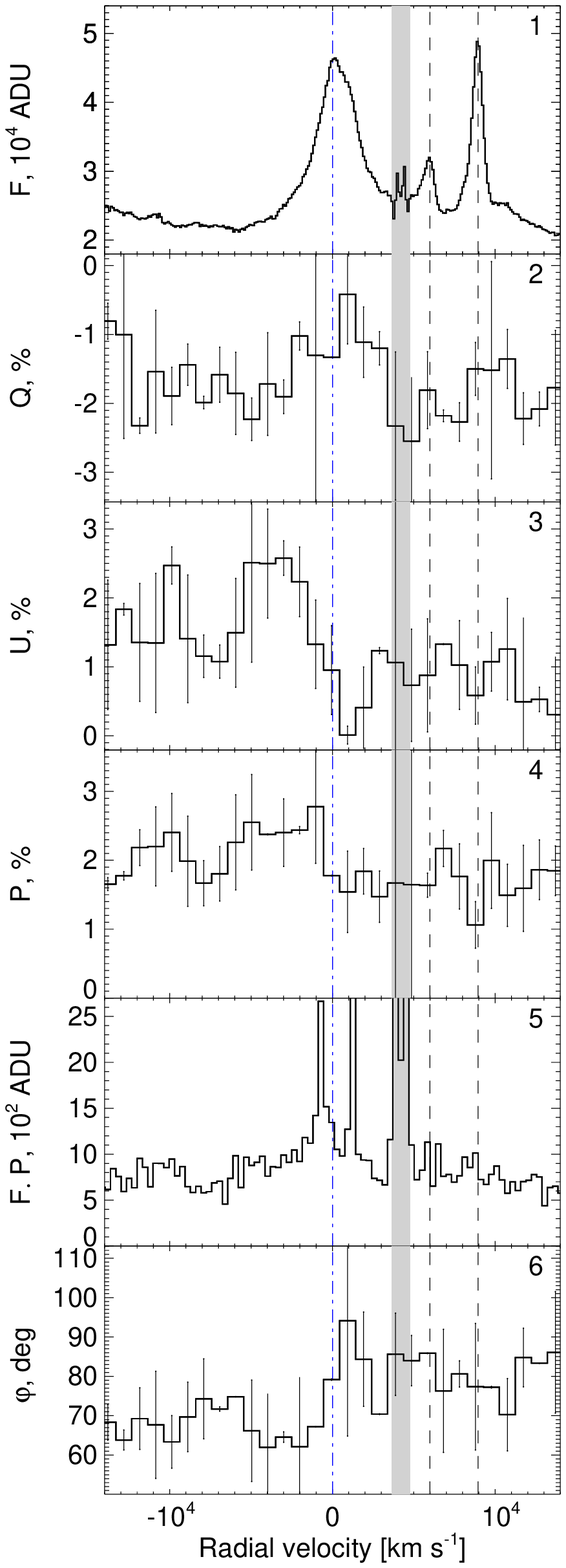}
     \end{subfigure}
     \begin{subfigure}[b]{0.4\textwidth}
         \includegraphics[width=\textwidth]{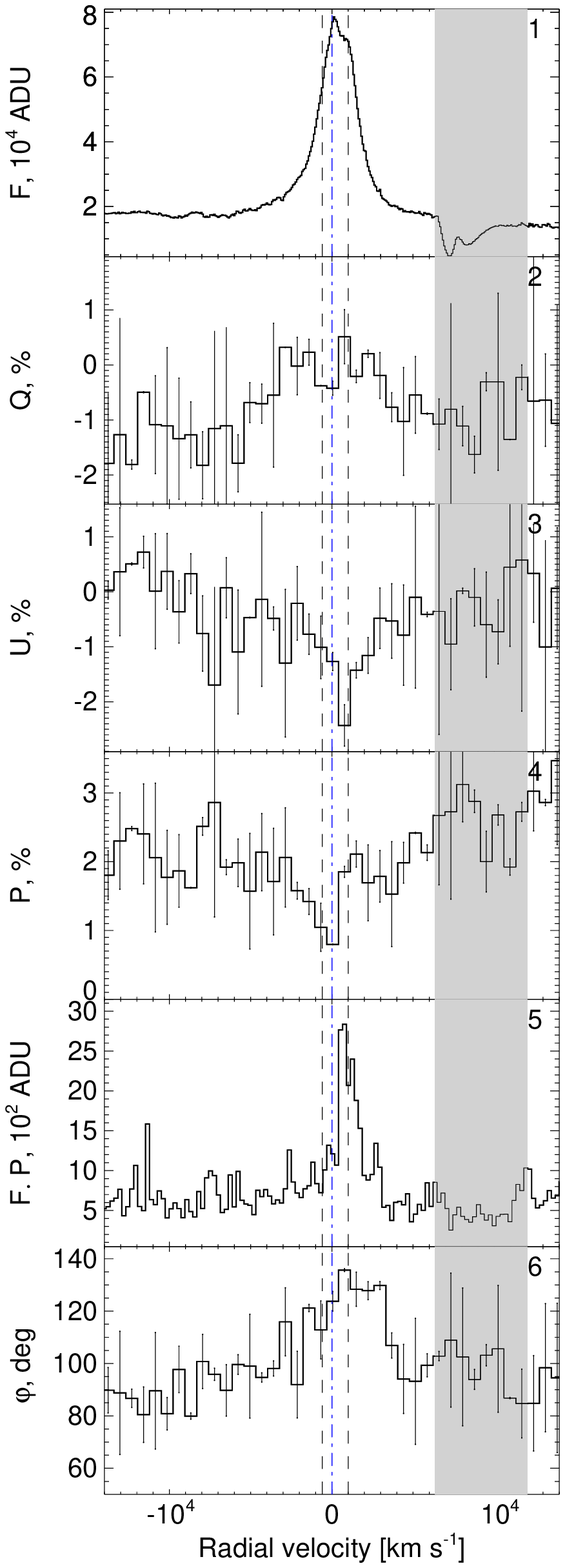}
     \end{subfigure}
        \caption{The profiles of H$\beta$ (\textbf{left}) and H$\alpha$ (\textbf{right}) lines. From~top to bottom: the intensity in the natural light, the~$Q$ and $U$ Stokes parameters, the~polarization degree, the~polarized flux, and the polarization angle. \xreplaced{Note here that the polarization degree is debiased according to the Equation~(\ref{formula:unbias}).}{} The values on panels 1 and 5 are binned in the 6 \AA\ window and the values on panels 2--4 and 6---in the 18 \AA\ window. The~black vertical dashed lines mark the narrow lines coming from the NLR---[OIII] lines at H$\beta$ profile and [NII] lines at H$\alpha$ profile. The~blue vertical dashed line marks the position of zero velocity. The~atmospheric absorption/emission bands are noted with light grey~stripes. }
        \label{fig:Zw_prof}
\end{figure}

Since the further analysis of the polarized spectrum of the object will concern only the profiles of two broad lines, Figure~\ref{fig:Zw_prof} presents spectral data H$\alpha$ and H$\beta$, recalculated in the velocity range from $-$14,000 to 14,000 km/s relative to the centers of the lines. Panels 1 and 5 demonstrate the flux $F$ and the polarized flux $F\cdot P$ in counts; the Stokes parameters $Q$ and $U$ (panels 2 and 3) are corrected for the polarization of the field stars calculated from the polarimetry data in the filters (the correction is equal to 3.1\% for $Q$ and $-$1.8\% for $U$). The~polarization degree $P$ and angle $\varphi$ are recalculated from the Stokes parameters $Q$ and $U$, taking into account the \xreplaced{bias}{} correction \xreplaced{(the Rice factor is about 0.67)}{}. For~the larger visibility the parameters $Q$, $U$, $P$ and $\varphi$ are binned in the window 18 \AA; the spectrum $F$ is given with 2 \AA\ resolution, the~polarized spectrum $F \cdot P$---with 6\AA\ resolution.

\subsubsection{Mass~Estimation}
\label{mass_est}

\begin{figure}[H]
     \begin{subfigure}[b]{0.45\textwidth}
         \includegraphics[width=\textwidth]{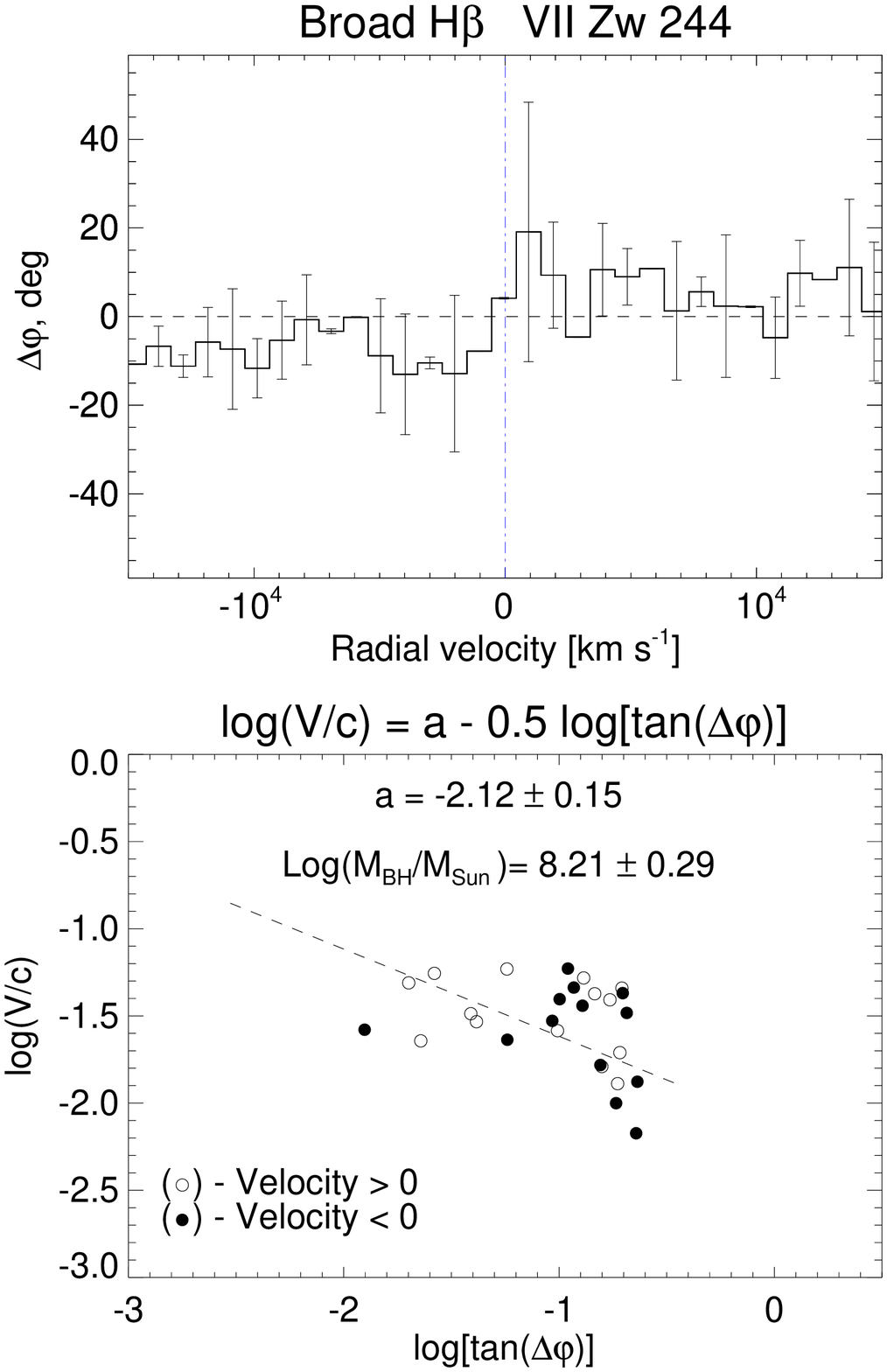}
     \end{subfigure}
     \begin{subfigure}[b]{0.45\textwidth}
         \includegraphics[width=\textwidth]{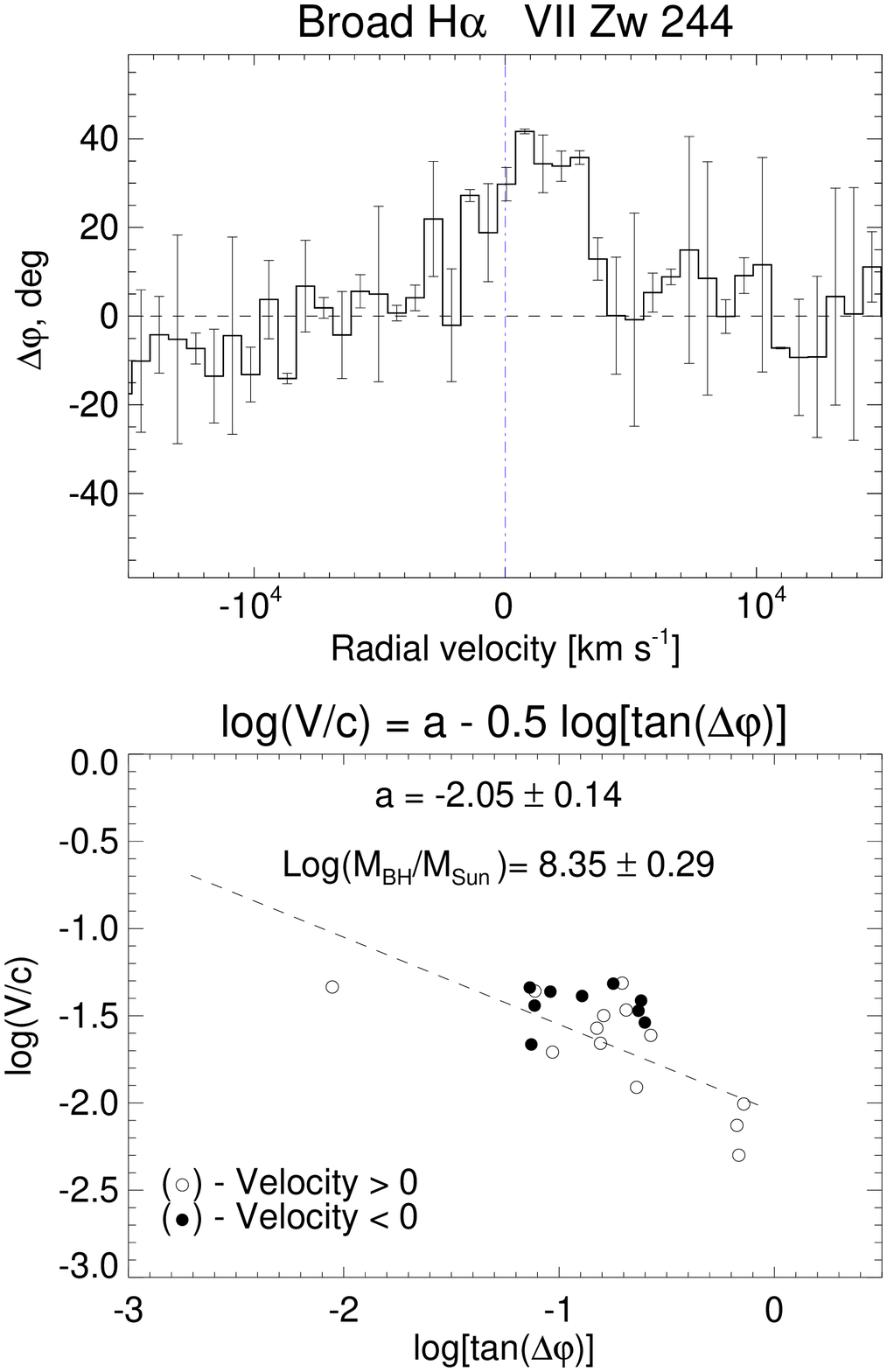}
     \end{subfigure}
        \caption{The polarization angle profiles for H$\beta$ (\textbf{left}) and H$\alpha$ (\textbf{right}) lines. On~the upper panels the deviation of the polarization angle to the mean value (horizontal dashed line) is given. \xreplaced{}{Blue lines mark the Keplerian motion.} The bottom panels show the relation of $\log[\tan(\Delta \phi)]$ v.s. $\log(V/c)$. }
        \label{fig:vel_dia}
\end{figure}

As shown by \citet{afanasiev15} and \citet{afanasiev19}, spectropolarimetric observations of galaxies with equatorial scattering make it possible to estimate the SMBH masses regardless of the inclination angle of the system to the observer. The~radiation of the differentially rotating disc-shaped BLR is reflected from the equatorial scattering region so that the observer registers the specific features in broad lines in polarized light, in~particular, the~S-shaped profile of the polarization angle, which is a marker of Keplerian motion in the disk. The~change of the polarization angle along the broad line is related to the gas rotation velocity in BLR:
\begin{equation}
    \log(V/c) = {\rm a} - 0.5\log[\tan(\Delta \varphi)],
\end{equation}
where $\Delta \varphi$ is relative variations of the polarization angle (see Figure~\ref{fig:vel_dia}), and~ $-$0.5 coefficient is consistent with the assumption of BLR Keplerian motion. The~parameter $a$ is a function of $M_{\rm BH}$ and can be written as
\begin{equation}
{\rm a}=0.5\log[\frac{GM_{{\rm BH}}\cos^{2}(\theta)}{c^{2}R_{{\rm sc}}}],
\end{equation}
where $R_{\rm sc}$ is the inner radius of the scattering region, and~$\theta$ is the angle between the BLR and the scatterer plane. From~the assumption of coplanarity, we assume $\theta = 0$, which contributes no more than $\sim10\%$\ uncertainty to the mass estimate. Then the SMBH mass can be expressed as follows:
\begin{equation}
\label{eq-mass}
\log(M_{\rm BH}/M_{\odot}) \approx 10.25 + 2{\rm a} + \log(R_{\rm sc})
\end{equation}

In Equation~(\ref{eq-mass}), the~parameter $a$ is determined from observational data, while the size of the scattering region $R_{\rm sc}$ is determined by indirect methods. It is commonly assumed that $R_{\rm sc}$ characterizes the inner boundary of the gas-dust region (so-called dusty torus). Currently, two methods based on IR observations are used to determine the \xdeleted{torus} size of the torus: interferometry in the near-IR and mid-IR range \cite{gravity2020} and IR reverberation mapping \cite{yang2020}.

\citet{lyu2019} presented the results of the reverberation mapping of the dust region of the galaxy VII Zw 244 in WISE W1 ($\sim$3.4 $\upmu$m) and W2 ($\sim$4.5 $\upmu$m) bands. The~IR radiation shows a delay of about 340--360 days relative to the optical emission, which indicates the size of the region emitting in the IR range.
However, IR observational methods localize deep regions of the dusty torus, where the optical thickness is $\gg$1, while equatorial scattering occurs in a region where the optical thickness is of the order of 1 and electron scattering dominates \citep{savic18}. An~alternative method has recently been proposed for determining the position of the equatorial scattering region based on the reverberation mapping in polarized broad lines \citep{shablovinskaya2020}. As~well as other reverberation methods, this approach requires a large amount of observational time, which is why it was still not applied to this galaxy. However,  \xreplaced{even the first}{} result for the Mrk 6 galaxy obtained \xreplaced{within the given approach}{} showed that the measured $R_{\rm sc}$ turned out to be 2 times smaller than the size of the dusty torus estimated in the IR band. This \xreplaced{case}{} allows us to assume that for VII Zw 244 $R_{\rm sc} \approx 170$ lt~days.

Additionally, we are able to calculate the value of $R_{\rm sc}$ based on scale relation from \citep{afanasiev19}. According to this work, $R_{\rm sc} \simeq 5.1 R_{\rm BLR}$, where we take the estimate of the size of the area BLR $R_{\rm BLR}$ from \xdeleted{the article} \citet{malygin20}: $R_{\rm BLR}=30.7^{+2.1}_{-2.3}$ lt days. Then $R_{\rm sc} \simeq 157$ lt days. This value turns out to be approximately equal to half the size of the dust region in the IR range, and~then we will use it to estimate the SMBH~mass.

Since for VII Zw 244 the signs of equatorial scattering were detected in two broad lines, we present the profiles of the polarization angle switch for H$\alpha$ and H$\beta$ (Figure \ref{fig:vel_dia}). The~upper panels show variations of the polarization angle relative to the average level. In~both lines, it can be seen that $\Delta\varphi$ to the right from the line center takes positive values; at the same time, the~left negative wing of the profile is detected only in the H$\beta$ line. \xreplaced{While the H$\alpha$ polarization angle profile is not symmetric as it is predicted by the model of the pure Keplerian motion, based on the sample provided in \citep{afanasiev19} one could find close examples of asymmetric polarization angle switch, as~e.g.,~Mrk 1501 (see Figure~\ref{fig:vel_dia} there).}{} It can be assumed that the blue wing of the H$\alpha$ line profile suffers from depolarization, which may be associated with radial outflows from the region \citep[]{savic2020}.

On the bottom panels of Figure~\ref{fig:vel_dia} the dependence of the logarithm of the gas velocity in the line $\log(V/c)$  on the logarithm of the \xreplaced{tangent }{tagent} of the polarization angle variations  $\log[\tan(\Delta\varphi)]$ is presented. \xreplaced{Regardless of the H$\alpha$ line profile asymmetry, the~data points of the measured values of the gas velocity $\log(V/c)$ show clearly the linear dependence on $\log[\tan(\Delta\varphi)]$ in both emission lines, which is a significant proof of the observed Keplerian-like motion.}{}  Fitting the observed points by linear regression we estimated ${\rm a}=-2.12 \pm 0.15$ for H$\beta$ and $-2.05 \pm 0.14$ for H$\alpha$. In~the case of both broad lines, the~mass estimates turn out to be close: $\log(M_{\rm BH}/M_{\odot}) = 8.21 \pm 0.29$ for H$\beta$ and $8.35 \pm 0.29$ for H$\alpha$. Note that if we assume $R_{\rm sc} =350$ lt days, as~follows from the IR observations, then $\log(M_{\rm BH}/M_{\odot}) \approx 8.6$, which is consistent with the estimate obtained above, within~the error. \xreplaced{Additionally, it is important to mention that the uncertainty obtained here was calculated as the variance of the deviation of the measured values of the gas velocity in the broad line from the Keplerian ones, which is assumed to be a random value distribution. As~it can be seen in \citep{afanasiev19}, 0.2--0.3 dex is a typical uncertainty for the spectropolarimetric mass measurements.}{}

\subsubsection{Bolometric~Luminosity}

For this object we are also using results of our observations on BTA 6 m telescope $\log{[L_\text{5100}(erg/s)]} = 44.22$ \citep{malygin20} and bolometric correction $BC = 10.3$. So $\log{[L_\text{bol}(erg/s)]} = 45.24$. Note that in \citet{lani17} a close value of $\log{[L_\text{5100}(erg/s)]} = 44.16$ was~obtained.

\subsubsection{Inclination Angle and Spin~Value} \label{inclinZw}

\xreplaced{}{\mbox{\citet{zhuang18}} obtained the value $i \approx 15^{\circ}$ for this object by quantifying the contribution of the torus luminosity to the total IR emission. }

As shown in \citep{afanasiev19}, the~inclination angle of the AGN can be obtained by comparing two independent measurements of the SMBH mass. The~estimate of the SMBH mass obtained from the reverberation mapping is known up to the dimensionless parameter $f$ \xreplaced{(see \linebreak Equation (\ref{vp})).}{\mbox{(see \cite{peterson2014})}:}
\xreplaced{}{
$   M_{\rm BH} = f \times (R_{\rm BLR}\vartheta^2_{\rm line}G^{-1}) = f \times V.P.,$  }
\xreplaced{}{where $\vartheta_{\rm line}$ is the velocity of the line-emitting gas in the BLR, $G$ is the gravitational constant, and~$V.P.$ is a virial product.}
In the paper by \mbox{\citet{malygin20}} the obtained value of the SMBH mass is given for $f=1$, so for VII Zw 244 $\log(V.P./M_\odot) = 7.05^{+0.07}_{-0.08}$. Comparing $V.P.$ with the mass estimate $\log(M_\text{BH}/M_{\odot}) = 8.3\pm 0.3$ obtained from spectropolarimetric approach, we get $f = 21.1\pm 10.1$. According to \mbox{\citet{yu2019}}, the~factor $f$ is connected to kinematics, geometry and inclination of the BLR clouds and can be determined as $f = [h^2+\sin^2(i)]^{-1},$ where $h$ is the ratio of the height to radius of the thick disk BLR (or the ratio of the turbulent velocity to the local Keplerian velocity for a given radius), which can be assumed $h=0$ \xreplaced{for geometrically thin disks}{}. Then for VII Zw 244 the angle of inclination is $i = (14.3 \pm 3.6)^{\circ}$. \xreplaced{Note here}{Additionally, note} that although the average inclination angle of AGNs with equatorial scattering is $\sim$35$^{\circ}$, objects with $i\sim 14^{\circ}$ have already been previously detected in the \xreplaced{sample of AGNs with equatorial scattering }{AGN sample}, e.g.,~Mrk 1148 \citep{afanasiev19}.

Using the same approach as for the previous one \citep{piotrovich22} and mass from \citet{malygin20} we get that $i \gtrsim 18^{\circ}$. \xreplaced{Moreover, \citet{zhuang18} obtained the value $i \approx 15^{\circ}$ for this object independently by quantifying the contribution of the torus luminosity to the total IR emission. }{} So, taking into account other estimates we concluded that in this case $i \approx 18^{\circ}$. As~can be seen, this value agrees well with the \xreplaced{independent estimates shown above obtained by IR SED analysis \citep{zhuang18} and SMBH masses estimates comparison }{previously obtained independent estimates}.  The~results of estimations of spin are presented in Table~\ref{tab01}. $a \approx 0.996$ is a rather typical value for objects of this~type.

In \citet{Huetal21} the authors obtained close values  of time lag (coinciding within the error limits). To~calculate $V.P.$, they measured the velocity by $\sigma_{line}$ (square root of the second moment) using the rms spectrum. The~authors obtained values of the virial product that are smaller than~in \citet{malygin20}, where the gas velocity was measured via the $\sigma_{line}$ from the single epoch spectrum of a good signal-to-noise ratio with decomposition and subtraction of narrow components near the H$\beta$ line. Thus, the~$V.P.$ from \citep{Huetal21}, compared with the spectropolarimetric mass measurement, gives \xreplaced{an}{a} inclination angle of the system of about $\sim$ 11$^{\circ}$, from~which one can conclude that their measurement of the virial product is probably~underestimated.

\subsubsection{Magnetic~Field}

\xreplaced{This object is radio quiet \citep{laor08}, however, according to recent studies \citep{berton20,silpa22,hartley19}, radio-quiet AGNs can have jets and emission mechanisms similar to radio-loud objects. Therefore, we decided that it would be useful to try to apply method from \citet{piotrovich20} \linebreak (Equation (\ref{eq02})) to this object as well. }{We estimated magnetic field strength using Equation~(\ref{eq02}).} For this object we also take value $\vartheta_\text{line}$ from \citet{malygin20} and calculate $FWHM = 2.335 \cdot \vartheta_\text{line}$ because in this case $\vartheta_\text{line}$ referred to H$\beta$ line.

This object has equatorial scattering, which does not allow us to consider the accretion disk as the only source of polarization. Thus, it is not possible to determine the magnetic field strength using the method from \citet{piotrovich21}.

The result is shown in Table~\ref{tab01}. The obtained value of the magnetic field is also quite~typical.


\section{Conclusions}\label{sec4}

Here we presented the results of the spectropolarimetric observations of 2 type 1 AGNs from the reverberation mapping sample from \citep{Uklein19}. The~measured polarization properties have been used for exploring polarization mechanisms and the determination of physical parameters of the given AGN. We outline the following conclusions:

\begin{itemize}
    \item for LEDA 3095839, no signs of equatorial scattering were detected in the polarized spectrum. At~the same time, we estimated the level of continuous polarization as $(0.9\pm 0.4)$\%, which is generated, as~we believe, in~an accretion disk. Using the approach from \citet{du14} we came to the conclusion that for this object it makes sense to consider two options: inclination angle $i \approx 35^\circ$ and $i \approx 45^\circ$. Thus, we obtained spin values $a = 0.966^{+0.030}_{-0.106}$ and $a = 0.736^{+0.226}_{-0.368}$ and magnetic field strengths at event horizon of SMBH $\log(B_\text{H}[G]) = 4.06^{+0.24}_{-0.24}$ and $\log(B_\text{H}[G]) = 4.00^{+0.56}_{-0.34}$ respectively. Additionally, we estimated magnetic field strength and the exponent $s$ of the power-law dependence of the magnetic field on the radius in accretion disk via polarimetric data using model from \citet{piotrovich21} for this two cases: $\log(B^*_\text{H}[G]) = 3.53^{+0.26}_{-0.53}$, $s = 1.77 \pm 0.18$ and $\log(B^*_\text{H}[G]) = 4.08^{+0.30}_{-1.08}$, $s = 1.63 \pm 0.23$.
    \item for VII Zw 244 we detected the specific features of equatorial scattering in both H$\alpha$ and H$\beta$ broad lines. It allowed us to estimate the SMBH mass as $\log(M_\text{BH}/M_{\odot}) = 8.3\pm 0.3$. The~comparison of the given mass with the previously obtained results gave the estimation of the inclination angle $i = (14.3 \pm 3.6)^{\circ}$, which allowed us to use the approach from \citet{du14} (estimation via the AGN luminosity)  assuming that $i \approx 18^\circ$ (minimum possible value for this object in this method). This approach give us the spin value $a = 0.996^{+0.002}_{-0.012}$ and magnetic field strength at event horizon of SMBH $\log(B_\text{H}[G]) = 4.29^{+0.10}_{-0.13}$.
\end{itemize}

\vspace{6pt}



\pagebreak
\authorcontributions{Conceptualization, M.P. and E.S.; methodology, M.P. and E.S.; software, M.P., E.S. and E.M.; validation, S.B. and T.N.; formal analysis, M.P.; investigation, M.P., E.S., E.M., S.B. and T.N.; resources, E.S. and E.M.; data curation, E.S. and E.M.; writing—original draft preparation, M.P., E.S. and E.M.; writing—review and editing, M.P., E.S. and E.M.; visualization, E.S. and E.M.; supervision, M.P. and E.S.; project administration, M.P. and E.S.; funding acquisition, E.S. All authors have read and agreed to the published version of the manuscript.}

\funding{This work was supported by the Russian Scientific Foundation (grant no. 20-12-00030 “Investigation of geometry and kinematics of ionized gas in active galactic nuclei by \linebreak polarimetry methods”).}

\dataavailability{The data underlying this article are available in the article.}

\acknowledgments{Observations with the SAO RAS telescopes are supported by the Ministry of Science and Higher Education of the Russian Federation. The~renovation of telescope equipment is currently provided within the national project "Science and~Universities". The authors are grateful to the reviewers for useful comments.}

\conflictsofinterest{The authors declare no conflict of~interest.}




\begin{adjustwidth}{-\extralength}{0cm}

\printendnotes[custom]

\reftitle{References}

\end{adjustwidth}
\end{document}